\documentclass[11pt]{article}
\usepackage{amsmath,amsbsy,amsfonts,amssymb,graphics,epsfig,float,mathrsfs,amsthm,braket}
\usepackage{ifsym}
\usepackage{psfrag}
\usepackage{color}
\usepackage[normalem]{ulem}

\definecolor{orange}{RGB}{255,127,0}
\definecolor{yellow}{RGB}{255,255,0}

\newcommand{\blue}[1]{{\color{blue} #1}} 
\newcommand{\red}[1]{{\color{red} #1}}    
\newcommand{\green}[1]{{\color{green} #1}}    

\newcommand \tR{\tilde{R}}
\newcommand \tZ{\tilde{Z}}
\newcommand \tPsi{\tilde{\Psi}}

\newcommand \epsM{\epsilon_m}
\newcommand \epsG{\epsilon_g}
\newcommand \beq{\begin{equation}}
\newcommand \eeq{\end{equation}}
\newcommand \beqn{\begin{equation*}}
\newcommand \eeqn{\end{equation*}}
\newcommand{\upd}{\mathrm{d}}
\newcommand{\pd}{\partial}
\newcommand{\srr}{\sigma_{rr}}
\newcommand{\sqq}{\sigma_{\theta\theta}}
\newcommand{\cF}{{\cal F}}
\renewcommand{\lg}{\ell_g}
\newcommand{\lc}{\ell_c}
\newcommand{\lcurv}{\ell_\ast}
\newcommand{\glv}{\gamma_{lv}}
\newcommand{\wo}{\delta}
\newcommand{\woc}{\delta_c}

\newcommand{\Wnought}{Z_{(0)}}
\newcommand{\Wone}{Z_{(1)}}
\newcommand{\Chinought}{\Psi_{(0)}}
\newcommand{\Chione}{\Psi_{(1)}}
\newcommand{\nonset}{n_{\mathrm{onset}}}

\newcommand{\atanh}{\mathrm{arctanh}}
\newcommand{\lm}{\lambda_-}
\newcommand{\lp}{\lambda_+}

\newcommand{\dPsi}{\mbox{$\tilde{\Psi}$}}
\newcommand{\dW}{\mbox{$\tilde{Z}$}}

\newcommand{\s}[1]{{\textsf{\textbf{#1}}}}

\begin{document}

\title{\s{Indentation of a floating elastic sheet: Geometry versus applied tension}}
\author{ \textsf{Finn Box$^{1,2}$, Dominic Vella$^2$, Robert Style$^3$ and Jerome A. Neufeld$^{1,4,5}$}\\ 
{\it$^{1}$ BP Institute, University of Cambridge, Cambridge, CB3 0EZ, UK}\\
{\it$^2$ Mathematical Institute, University of Oxford, OX2 6GG, UK}\\
{\it$^3$ Department of Materials, ETH Z\"{u}rich, 8093 Z\"{u}rich, Switzerland}\\
{\it$^4$ Bullard Laboratories, Department of Earth Sciences, University of Cambridge,}\\
{\it Cambridge, CB3 0EZ, UK}\\
{\it$^5$ Department of Applied Mathematics and Theoretical Physics,}\\
{\it University of Cambridge, Cambridge, CB3 0WA, UK}\\
}

\date{\today}

\maketitle

\hrule\vskip 6pt
\begin{abstract}
The localized loading of an elastic sheet floating on a liquid bath occurs at scales from a frog sitting on a lily pad to a volcano supported by the Earth's tectonic plates. The load is supported by a combination of the  stresses within the sheet (which may include applied tensions from, for example, surface tension) and the hydrostatic pressure in the liquid. At the same time, the sheet deforms, and may wrinkle, because of the load. We study this problem in terms of the (relatively weak) applied tension and the indentation depth. For small indentation depths, we find that the force--indentation curve is linear with a stiffness that we characterize in terms of the applied tension and bending stiffness of the sheet. At larger indentations the force--indentation curve becomes nonlinear and the sheet is subject to a wrinkling instability. We study this wrinkling instability close to the buckling threshold and calculate both the number of wrinkles at onset and the indentation depth at onset, comparing our theoretical results with experiments. Finally, we contrast our results with those previously reported for very thin, highly bendable membranes.
\end{abstract}
\vskip 6pt
\hrule

 \maketitle

\section{Introduction}

Poking is a natural way in which to test the material properties of an object, both in everyday life (for example an under-inflated bicycle tyre) or, more quantitatively, in AFM measurements of graphene \cite{Lee2008,LopezPolin2015} and biological cells \cite{Pelling2004}. While in many situations, the object being poked is a homogeneous bulk material, in others the object is a composite, consisting, for example, of a bulk material with  a thin coating. In such scenarios, poking may provide information about the coating, the substrate that is coated or some combination of the two. 

The canonical problem to understand the relative roles of coating and substrate is that of a thin elastic film bonded to a substrate. Perhaps the simplest substrate response is one that provides a restoring force linear in the vertical deflection --- a Winkler foundation \cite{Hetenyi1946}. Physically, this corresponds to an object floating on the surface of a liquid: the hydrostatic pressure within the liquid provides a restoring force that is precisely linear in the vertical deflection. However, this linear response is also commonly used as a model of an elastic substrate --- this model assumes that the substrate consists of an array of linear springs and is therefore also known as the mattress model.

At the same time as being relatively simple to formulate mathematically, this scenario is also of interest at a range of scales: at very large scales, floating ice sheets are often used in cold regions as construction platforms for transport routes, airfields and offshore oil exploration sites. Determining the bearing capacity and failure of ice sheets subject to vertical loads is essential when assessing the operational potential of floating ice sheets \cite{Sodhi1995,Sodhi1998,Beltaos2002,Bazant2002}. This requires knowledge of the bending rigidity of sea ice \cite{Fox2001,Fox2000}, which is usually measured by comparison with theoretical results for the loading of a thin floating plate \cite{Kerr1976,Nevel1977}. At still larger scales, the loading of ice sheets by surface melt water has been implicated in the catastrophic collapse of the Larsen B ice sheet in Antarctica~\cite{Banwell2013} while at global scales the  gravitational loading of the lithosphere by mountain ranges \cite{Karner1983,Mahadevan2010} and volcanic sea mounts~\cite{Asaadi2011} involve much the same physical ingredients. 

At the other end of the length-scale spectrum, the elastic properties of thin biological materials may be characterized by measuring the deflection that results from an applied central point force \cite{Liu2001,Ju2002,Ahearne2007}. Similarly, the material properties of ultra-thin polymer films can be determined from the readily observable wrinkle patterns that form when floating films are subject to a localized force either from  the capillary pressure of a fluid droplet or an imposed displacement from an indenter \cite{Huang2007,Schroll2013,Vella2015}. In both cases, a vertical deflection pulls material radially inwards and in so doing generates a compressive azimuthal stress in the film that ultimately results in a radial pattern of wrinkles. The properties of these wrinkling patterns at very small scales have been extensively studied both  `near-threshold' (close to the onset of instability) \cite{Adams1993} and  `far-from-threshold' (once the wrinkling pattern is well-developed) \cite{Davidovitch2011,Davidovitch2012,Schroll2013,Vella2015,Paulsen2016}. The key observation is that in  highly flexible films, the stress state is qualitatively changed by wrinkling: wrinkling relaxes the stress in the direction perpendicular to the wrinkles \cite{Pipkin1986,Steigmann1990,Davidovitch2011}. This may have important consequences for the mean shape of the wrinkled object, which is, in general, different  to what would be observed in the absence of wrinkles \cite{Vella2015,Vella2015epl}. Furthermore, this wrinkling can have the surprising consequence that the force--displacement response depends not on the mechanical properties of the film (its modulus and thickness), but rather only its geometry (e.g.~radius) and other physics in the system \cite{Vella2015,Vella2015epl}. As well as their aesthetic appeal, these wrinkle patterns are of interest as a means of generating surfaces with functional patterned topology \cite{Chung2010} that may be useful in applications such as wetting \cite{Chen2016} and photonic devices \cite{Kim2012}. 


Although the large and small length scale problems discussed above contain the same physical ingredients, the former are dominated by bending stresses and gravity, while the latter are dominated by the surface tension of the interface, together with gravity. Indeed, it is this influence of the liquid surface tension that distinguishes large scales from small scales. The two problems may therefore be thought of as two limits of an elastic sheet floating at the surface of a liquid and subject to a tension at its boundary. Here we study the relative effects of the sheet's bending stiffness and the applied (interfacial) tension, focussing, in particular, on the transition between regimes in which one dominates the other. However, we shall also see that a key third ingredient is the amount of imposed deformation.

The paper is structured as follows: the detailed mathematical model used to describe this system is discussed in \S2.  The experimental set up used to study axisymmetric deformations of the floating sheet is described in \S3. In \S4, axisymmetric deformations are considered theoretically with the results of numerical and analytical arguments compared with experiments. This section finishes with a discussion of what distinguishes `small' and `large' deflections. The onset of wrinkling is detailed in \S5, including a linear stability analysis of the axisymmetric state, a description of the experimental technique employed to identify the onset of wrinkling and comparison between experimental results and the linear stability analysis,  before conclusions are presented in \S6. 

\section{Theoretical setting}\label{sec:theory}

\begin{figure}
\centering
\includegraphics[width = 0.75\textwidth]{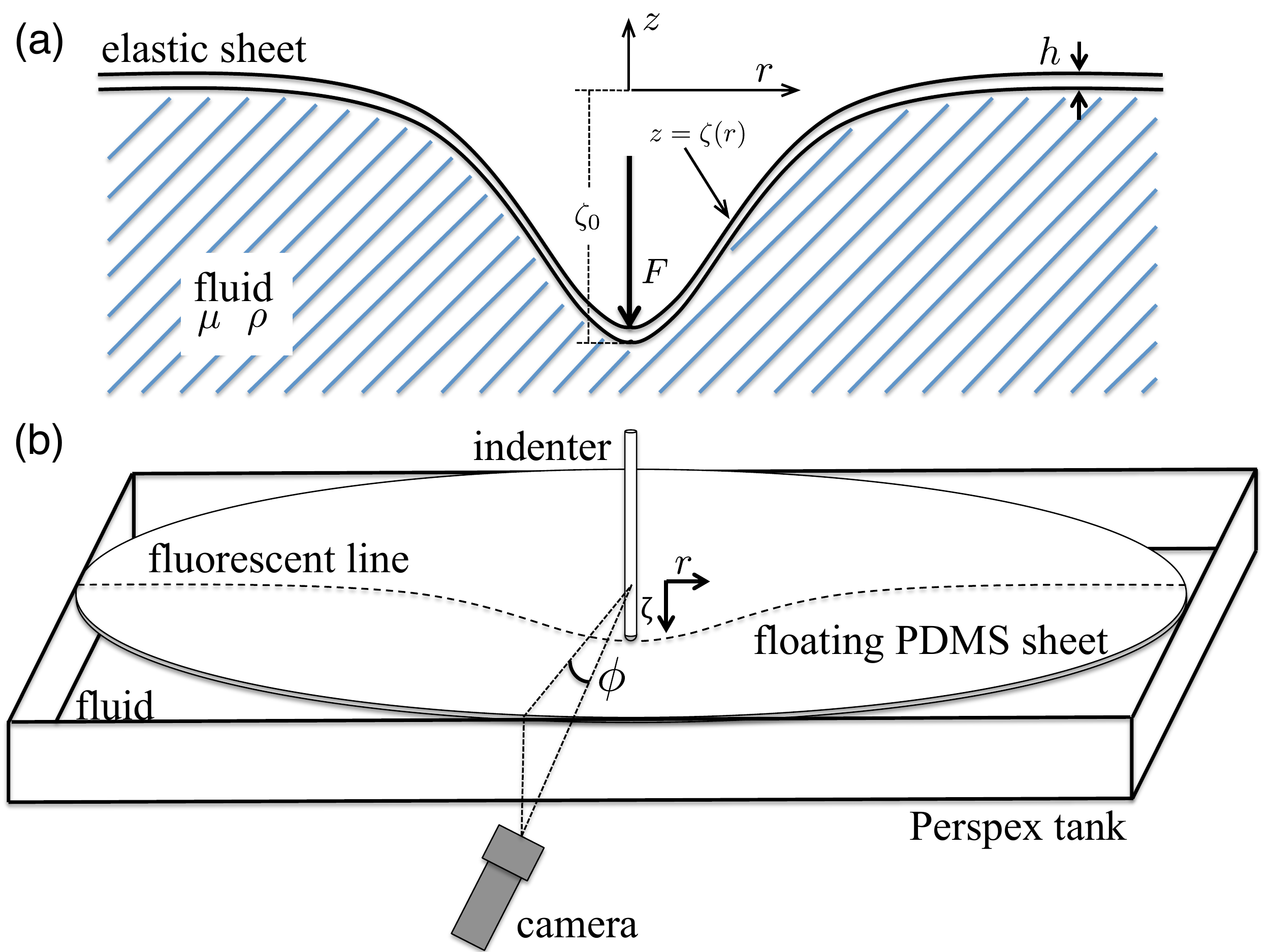}
\caption[Schematic diagrams]{(Color online). Schematic diagrams of the (a) model system and (b) experimental set up. In both cases, the vertical displacements of the sheet are measured from the free-floating equilibrium of the sheet.\label{fig:Schematic}}
\end{figure}

We consider an elastic sheet of thickness $h$, Young's modulus $E$ and Poisson ratio $\nu$, floating on a fluid of density $\rho$. The sheet is subject to a point-like force $F$ at its centre (shown schematically in figure~\ref{fig:Schematic}) which results in the deformation of the sheet.  Provided that deformations occur over a length scale that is large compared to the thickness of the sheet, we may model the resulting elastic deformation using the  F\"{o}ppl-von-K\'{a}rm\'{a}n equations, incorporating the hydrostatic pressure exerted by the fluid phase on the elastic sheet. Accordingly, the vertical displacement of the sheet from its neutral floating equilibrium, $\zeta(r,\theta)$, satisfies the vertical force balance equation \cite{Vella2015},
\begin{equation}
B\nabla^4\zeta-[\zeta,\chi]=-\rho g \zeta-\frac{F}{2\pi}\frac{\delta(r)}{r},
\label{eqn:fvk1dim}
\end{equation} where $B=Eh^3/[12(1-\nu^2)]$ is the bending stiffness of the sheet,  $g$ is the acceleration due to gravity, $\delta(r)$ is the Dirac $\delta$-function and
the operator $[f,g]$ is given in polar coordinates by \cite{Ventsel2001},
\beq
[f,g]=\frac{\pd^2f}{\pd r^2}\left(\frac{1}{r}\frac{\pd g}{\pd r}+\frac{1}{r^2}\frac{\pd^2 g}{\pd \theta^2}\right)+\frac{\pd^2g}{\pd r^2}\left(\frac{1}{r}\frac{\pd f}{\pd r}+\frac{1}{r^2}\frac{\pd^2 f}{\pd \theta^2}\right)-2\frac{\pd}{\pd r}\left(\frac{1}{r}\frac{\pd f}{\pd \theta}\right)\frac{\pd}{\pd r}\left(\frac{1}{r}\frac{\pd g}{\pd \theta}\right).
\eeq

In \eqref{eqn:fvk1dim}, the Airy stress function $\chi(r,\theta)$ is a potential for the in-plane stress and is introduced to ensure that the stress within the solid sheet automatically satisfies the equilibrium equation; this is achieved by setting $$\sqq=\frac{\pd^2\chi}{\pd r^2},\quad \srr=\frac{1}{r}\frac{\pd\chi}{\pd r}+\frac{1}{r^2}\frac{\pd^2\chi}{\pd \theta^2},\quad \mbox{and}\quad \sigma_{r\theta}=-\frac{\pd}{\pd r}\left(\frac{1}{r}\frac{\pd \chi}{\pd\theta}\right).$$  

A second differential equation arises from the ``compatibility of strains", the requirement that the strains associated with the stress of a particular stress function, $\chi(r,\theta)$, match the geometric strains associated with a particular out-of-plane displacement, $\zeta(r,\theta)$. This condition may be written
 \begin{equation}
\nabla^4\chi=-\frac{1}{2}Eh[\zeta,\zeta],
\label{eqn:fvk2dim}
\end{equation} 
where the product $Eh$ is the stretching stiffness of the sheet.

Indentation may be achieved by imposing a given force $F$ and measuring the indentation depth $\zeta_0$ that results, or by imposing an indentation depth and measuring the force required to produce this indentation. Both techniques were employed in our  experiments, described in \S\ref{sec:expt}, and are mathematically equivalent in terms of the model developed here. However, in numerical calculations, it is simpler to prescribe the indentation depth $\zeta(0)=\zeta_0$ and calculate the force $F$ that is required to achieve this level of indentation subsequently.  The conditions imposed on $\zeta$ at the indentation point are therefore the indentation depth, that the sheet does not have a cusp, and the requirement of zero horizontal displacement,
\beq
\zeta(0,\theta)=-\zeta_0,\quad \left.\frac{\pd \zeta}{\pd r}\right|_{(0,\theta)}=0, \quad u_r(r = 0) = 0.
\eeq 
The last condition on the horizontal displacement at the origin is equivalent to a condition on the hoop strain, $\lim_{r\to0}[r\epsilon_{\theta\theta}]=0$, which may, using Hooke's law be restated as a condition on the stress distribution at the origin, and hence on the derivatives of the Airy stress function $\chi$, namely
\beq
\label{eqn:dimstressbcsA}
\lim_{r\to0}\left[r\frac{\pd^2\chi}{\pd r^2}-\nu \frac{\pd\chi}{\pd r}-\nu\frac{1}{r}\frac{\pd^2\chi}{\pd \theta^2}\right]=0.
\eeq 
We also have an arbitrary choice of gauge in $\chi$ (since it is only the derivatives of $\chi$ that have physical relevance), and so we take
\beq
\label{eqn:dimstressbcsB}
\chi(0,\theta)=0.
\eeq

We consider a sheet where out-of-plane deformation occurs over a distance much smaller than the radial extent of the sheet. The effect of the finite radius of the sheet is known to play an important role in the development of wrinkle patterns in very thin sheets \cite{Vella2015}. Here we use larger sheets for which the effects of finite size are negligible (we shall discuss later what `large' means in this context). We therefore expect that, far from the indenter, the sheet will return to its undeformed, freely floating position, i.e.
\beq
\zeta,\frac{\pd\zeta}{\pd r}\to0 \quad\mbox{as}\quad r\to\infty.
\eeq 
In the far field we imagine that the sheet is subject to a homogeneous, isotropic tension which, in our experiments, arises from the surface tension of the liquid, $\glv$; we therefore have $\srr,\sqq\to\glv$ as $r\to\infty$, which requires that 
\beq
\chi\sim \glv \frac{r^2}{2} \quad \mbox{as} \quad r\to\infty.
\eeq 

\subsection{Non-dimensionalization}

In this paper, we seek to describe both relatively thick sheets (where we expect the applied, far-field tension due to surface tension to be a perturbative effect) and thinner sheets where this tension is important. We therefore choose a non-dimensionalization that does not break down in the limit $\glv\to0$. In particular, we let
\beq
Z=\zeta/h,\quad R=r/\lg,\quad \Psi=\chi/B,\quad \cF=F/(Bh/\lg^2)
\label{eqn:NonDim}
\eeq where
\beq
\lg=(B/\rho g)^{1/4}
\label{eqn:lgDefn}
\eeq is the \emph{elasto-gravitational} length scale over which the
deformation of the elastic sheet produces a bending stress comparable to the buoyancy force from 
 the fluid.

With this non-dimensionalization, we find that \eqref{eqn:fvk1dim} becomes
\beq
\nabla^4Z-[Z,\Psi]=-Z-\frac{\cF}{2\pi}\frac{\delta(R)}{R},
\label{eqn:fvk1nd}
\eeq while \eqref{eqn:fvk2dim} becomes
\beq
\nabla^4\Psi=-6(1-\nu^2)[Z,Z].
\label{eqn:fvk2nd}
\eeq Equations \eqref{eqn:fvk1nd}--\eqref{eqn:fvk2nd} are to be solved subject to the boundary conditions
\begin{equation}
Z(0)=-\wo=-\zeta_0/h,\quad Z'(0)=0,\quad \lim_{R\to0}\left[R\frac{\pd^2\Psi}{\partial R^2}-\nu \frac{\pd\Psi}{\pd R}-\nu\frac{1}{R}\frac{\pd^2\Psi}{\pd \theta^2}\right]=0, \quad \Psi(0)=0.
\end{equation} 
Far from the indenter we also have
\begin{equation}
Z,Z'\to0,\quad\Psi\to\tfrac{1}{2}\tau R^2 \quad \mbox{as} \quad R\to\infty.
\end{equation} Here
\beq
\tau=\frac{\glv}{(\rho gB)^{1/2}}
\label{eqn:tauDefn}
\eeq is the dimensionless 'applied tension'. We note that $\tau$ is the ratio of the relevant applied stress (surface tension, $\glv$) to the bending stresses when deformations occur on a length scale $\lg$, i.e.~$B/\lg^2$;  hence $\tau$ may also be thought of as a `mechanical bendability', $\epsM^{-1}$ in the terminology of Hohlfeld \& Davidovitch \cite{Hohlfeld2015}.     In our experiments, $\tau$ was varied predominantly by changing the thickness of the sheets, but also by changing the interfacial tension, to attain values in the range $10^{-3}\lesssim \tau\lesssim 30$. This range covers a wide range of behaviours and allows us to observe the beginning of the transition to extremely bendable, ultra-thin films with  $\tau\gtrsim10^4$ that have been studied previously \cite{Vella2015,Paulsen2016}. 

Another measure of the bendability of a thin elastic sheet exists, besides the mechanical bendability $\tau$: since indentation will itself induce a stress within the sheet, there is also a `geometrical bendability' \cite{Hohlfeld2015}. To determine this geometrical bendability, we note that if an indentation of amplitude $\zeta_0$ decays over a horizontal length scale $\lcurv$, then the geometry-induced stress is $Eh(\zeta_0/\lcurv)^2$, while the bending-induced stress is $B/\lcurv^2$. The geometrical bendability is therefore
\begin{equation}
\epsG^{-1} =  \frac{Eh\zeta_0^2}{B} \sim \left(\frac{\zeta_0}{h}\right)^2=\wo^2.
\end{equation} This geometrical bendability is simply the square of the dimensionless indentation depth, $\wo$, up to constants whose only dependence on the sheet's properties is through the Poisson ratio $\nu$.

Our problem is therefore governed by two dimensionless parameters: the geometrical and mechanical bendabilities, $\epsG^{-1}$ and $\epsM^{-1}$, respectively. While one might expect, on counting grounds, there to be another quantity measuring the significance of bending stresses to  the stretching stiffness of the sheet, i.e.~$q=(B\rho g)^{1/2}/(Eh)$, we note that this $q\propto (h/\lg)^2$, which must be small for our use of the F\"{o}ppl-von K\'{a}rm\'{a}n equations to be appropriate. Nevertheless, this `nearly inextensible' limit is a regular limit \cite{Vella2015}.

Focussing instead on the two parameters $\wo=\epsG^{-1/2}$ and $\tau=\epsM^{-1}$, we note that the geometrical bendability is independent of the bending stiffness of the sheet, liquid density and applied tension while the mechanical bendability depends on all of these and is, instead, independent of the imposed indentation. We also emphasize that the behaviour of the limits $\tau\ll1$ and $\tau\gg1$ are quite different. In the limit $\tau\to0$, the problem is perfectly regular, but only the parameter $\delta$ remains. As a result, we expect the behaviour of the sheet to be determined only by the value of $\delta$. A consequence of this is that features of the wrinkling instability in the sheet (e.g.~the critical indentation depth at which wrinkles appear) must be described by order one numbers as $\tau\to0$:  we can immediately see that the dimensionless critical indentation depth $\woc=O(1)$, so that the dimensional critical indentation depth $\zeta_0^{(c)}\propto h$, and the number of wrinkles at onset $\nonset=O(1)$ also. In contrast, when $\tau\gg1$ ($\epsM^{-1}\gg1$), both of the dimensionless parameters $\tau$ and $\wo$ will matter; in particular, we should expect to observe a dependence of $\woc$ and $\nonset$ on $\tau$. In table \ref{table:DG} we highlight the relevant dimensionless groups (DGs) observed in each of the regimes $\tau\ll1$ and $\tau\gg1$, as well as focussing on which features of the problem have been studied previously, are the focus of this work, or remain open problems for future work.

\begin{table}
\begin{center}
\begin{tabular}{ |c|c|c|c|c|c|}
\hline
& Relevant DGs &  $\delta_c$  &  $\nonset$ & Limit $\delta\ll\delta_c$ & Limit $\delta\gg\delta_c$  \\ 
\hline 
& & & & & \\
$\tau \ll 1$ &  $\delta=\epsG^{-1/2}$ & This work  & This work & \cite{Hertz1884} & Open  \\
&&&&& \\
$\tau \gg 1$ & $\delta=\epsG^{-1/2}$ &   \cite{Vella2015} & This work & This work &  \cite{Vella2015}   \\
&$\tau=\epsM^{-1}$&&&& \cite{Paulsen2016} \\
\hline
\end{tabular}
\caption[Dimensionless groups]{Dimensionless groups (DGs) that govern the behaviour of a floating elastic sheet subject to a dimensionless indentation (geometrical bendability) $\delta$ and applied tension (mechanical bendability) $\tau$. Aspects of the limit $\tau\gg1$ have been considered previously, particularly the critical indentation depth required for wrinkling, $\delta_c$, and the behaviour of the system far beyond this threshold. However, the detailed computations of the number of wrinkles at onset, $\nonset$, reported here is novel. For the limit $\tau\ll1$, very little has been studied previously. } 
\label{table:DG}
\end{center}
\end{table}

\section{Experimental measurements of axisymmetric deformation \label{sec:expt}}

A schematic diagram of the experimental apparatus used to measure the elastic response of a floating sheet to a localized load is shown in figure~\ref{fig:Schematic}(b). Initially, a series of experiments were performed on relatively large floating elastic sheets for which the mechanical bendability was small, $\tau \ll 1$. In this series of experiments, detailed in \S\ref{sec:expt}(a), sheets of varying thickness $h$ were indented with a known applied force and the resulting, axisymmetric, profiles of the sheets were measured. Further experiments were then performed on smaller, thinner sheets of varying bending stiffness $B$  to assess the influence of $\tau$. In these experiments, described in \S\ref{sec:expt}(b), the centre of the sheet was indented to a known displacement and the force required to achieve such a deformation was measured. 
Together, the results from both sets of experiments map the transition from a regime in which bending controls the axisymmetric deformation to one in which the in-plane tension dominates instead. The experimental techniques used to study the wrinkling that occurs for large-amplitude deformation are described separately, in \S5(b).

\subsection{Low mechanical bendability, $\tau \ll 1$ }
\label{subsec:thick}

A first series of experiments were performed using a range of circular sheets of Polydimethylsiloxane (PDMS) with diameter $D\gtrsim0.5\mathrm{~m}$ and thickness $h\gtrsim 1.5\mathrm{~mm}$. These sheets were produced by spreading a commercial silicone elastomer (Sylgard 184, Dow Corning, UK) on a carefully levelled table, and then curing the sample for one week in a temperature-controlled room that maintained the temperature in the range $40-45^{\circ}$\,C. The Young's modulus and Poisson ratio of these sheets were measured by performing compressive tests using an Instron 3345 and were found to be $E = 2.06\pm0.03$\,MPa and $\nu = 0.50\pm0.01$, respectively. The detailed properties of the PDMS sheets used in these experiments are given in table \ref{table}. 

For each experiment an elastic sheet was carefully positioned on top of a layer of water contained within a tank of square cross-section and area $50^2$\,cm$^{2}$ or $1$\,m$^{2}$, depending on the diameter of the sheet in use. The density of PDMS, $\rho_{\textrm{\tiny PDMS}} \approx 929$\,\mbox{kg m}$^{-3}$, is less than that of water at 20$^{\circ}$C, $\rho = 998$\,\mbox{kg m}$^{-3}$, so that the sheet floats in equilibrium. The edge of the elastic sheet was freely floating (with no normal force or bending moment applied). However, spacers attached to the internal tank walls reduced the size of the tank cross-section to that of the sheet diameter at four positions and contacted the sheet to ensure it did not rotate during experiments, whilst minimizing any effect on the stress within the sheet.

In these experiments the mechanical bendability was calculated to be $\tau\lesssim 10^{-2}$. The relative insignificance of surface tension was confirmed by adding surfactant (washing-up liquid) to the liquid bath. The addition of surfactant reduced the surface tension of the fluid from $\gamma_{lv} = 72.8\mathrm{~Nm^{-1}}$ to $\gamma_{lv} = 24.9\mathrm{~Nm^{-1}}$, as was measured using a Drop Shape Analyser (DSA100, Kr\H{u}ss GmbH, Germany). At the concentrations used the density of the water remains unaltered. Despite a reduction in surface tension of more than a factor of two, the results obtained here were quantitatively indistinguishable. Further, our surface tension measurements did not change when compared before and after the experiment: any free polymer chains released by the sheets did not modify the surface tension of our (relatively large bath) significantly, as recently reported for small droplets \cite{Hourlier2017}.

A localized force was applied to the centre of the floating sheet using an indenter of length $300$\,mm, diameter $6.0\pm0.05$\,mm with a hemispherical end cap (making contact with the sheet). The radius of this contacting cap, $r_{\mathrm{cap}}\approx3\mathrm{~mm}$ is significantly smaller than the relevant horizontal length scale ($\lg\gtrsim15\mathrm{~mm}$ throughout this series of experiments); we therefore expect the point indenter approximation to be reasonable, as we shall discuss in due course. The indenter was weighted to obtain a given applied force in the range $0.34 - 20.76$\,N with an accuracy of $\pm 0.001$\,N, and was held inside a guiding tube to ensure the application of a central, vertical force. 

For a small applied force, and therefore small  indentation depth, the vertical deflection of the sheet remained axisymmetric. The magnitude of this axisymmetric deformation was determined by digitally imaging the deflection of a line drawn along the bottom surface of the sheet. The line was $\sim2$ mm in width, fluorescent and illuminated using a blue-light LED lamp. The entire sheet was imaged using a Nikon D5000 with a resolution of $4288 \times 2848$ pixels which was positioned at 27$^{\circ}$ to the horizontal and perpendicular to the line.  A high-pass filter positioned between the line and imaging camera produced a high contrast image, enabling the line to be distinguished from other features in the experiment. The deflection of the line was measured with respect to a reference image of the undeflected line.  Vertical deflections were resolved to within $100\mathrm{~\mu m}$ by fitting a Gaussian profile across the line, and processing the differences between images of the deflected line and the reference image \cite{Lister2013}.  

\subsection{Moderate mechanical bendability, $\tau \gtrsim 10^{-1}$}

Further experiments were performed on thinner sheets for which the mechanical bendability was no longer small (in particular, $\tau\gtrsim0.1$). To obtain significantly thinner sheets of uniform thickness, we used spin coating of two grades of polyvinylsiloxane (PVS) elastomer (Elite 8 Double and Elite 22 Double, Zhermack, Italy).  The thickness of the spin-coated sheets was measured prior to experimentation using a microscope (Leica DMIL, Leitz Wetzlar, Germany).  The Young's modulus of the cured elastomers were measured by performing tensile tests on a structural testing machine (Instron 3345, Instron, UK) and found to be $200\pm 13$\,kPa and $801\pm18$\,kPa, respectively, with $\nu = 0.5$ in both cases\footnote{In Tables~\ref{table} and \ref{table2}, PVS$_8$ and PVS$_{22}$ correspond to elastomers Elite double 8 and 22, which had measured moduli $E = 200$\,kPa and $E = 801$\,kPa, respectively.}. Experiments were also conducted on a Polyimide (PI) sheet of $h=8.5\pm2.0$\,$\mu$m, $E = 3.7\pm0.7$\,GPa and $\nu = 0.34$ (supplied by Goodfellow, Cambridge). The material properties of the PVS and PI sheets are detailed in Table~\ref{table}.

Limitations of spin coating meant it was not possible to obtain very large, thin sheets. Here, we used circular sheets of diameter $D = 89\pm0.5$\,mm, floating on a layer of water contained within a petri dish of inner diameter $D = 91\pm0.5$\,mm.

The petri dish containing water and a floating thin sheet was itself positioned upon a microbalance (Pioneer PA64C Analytic Balance, Ohaus, Switzerland). The centre of the sheets were indented by a needle tip of diameter $0.4$\,mm attached to a linear actuator (M228, Physik Instrumente, Germany) that was driven by a computer-controlled stepper motor (Mercury Step C663, Physik Instrumente, Germany); reported vertical deflections were accurate to $0.1\mathrm{~\mu m}$. The applied force was measured by recording the mass reported by the microbalance (accurate to within $0.1\mathrm{~mg}$).

\section{Axisymmetric deformations\label{sec:axisymm}}

Having outlined the experimental techniques used for studying axisymmetric deformations, we now return to the theoretical setting discussed in \S\ref{sec:theory}, and specialize to the case of axisymmetric deformations,  i.e.~$Z(R,\theta)=Z(R)$. 

\subsection{Numerical solution}

For  axisymmetric deformations, the vertical force balance and compatibility equations, \eqref{eqn:fvk1nd} and \eqref{eqn:fvk2nd}, become a pair of coupled, nonlinear ordinary differential equations,
\beq
\frac{1}{R}\frac{\upd}{\upd R}\left\{R\frac{\upd}{\upd R}\left[\frac{1}{R}\frac{\upd}{\upd R}\left(R\frac{\upd Z}{\upd R}\right)\right]\right\}-\frac{1}{R}\frac{\upd}{\upd R}\left(\frac{\upd Z}{\upd R}\frac{\upd \Psi}{\upd R}\right)=-Z-\frac{\cF}{2\pi}\frac{\delta(R)}{R}
\label{eqn:fvk1ndaxis}
\eeq and
\beq
R\frac{\upd}{\upd R}\left[\frac{1}{R}\frac{\upd}{\upd R}\left(R\frac{\upd\Psi}{\upd R}\right)\right]=-6(1-\nu^2)\left(\frac{\upd Z}{\upd R}\right)^2,
\label{eqn:fvk2ndaxis}
\eeq subject to force and symmetry boundary conditions on the deflection at the origin,
\beq
Z(0)=-\wo,\quad Z'(0)=\lim_{R\to0}\bigl[R\Psi''-\nu \Psi' \bigr]=\Psi(0)=0,
\label{eqn:bc1axis}
\eeq and far-field conditions
\beq
Z,Z'\to0,\quad\Psi\to\tfrac{1}{2}\tau R^2 \quad\quad (R\to\infty).
\label{eqn:bc2axis}
\eeq

\begin{figure}
\centering
\includegraphics[width=\textwidth]{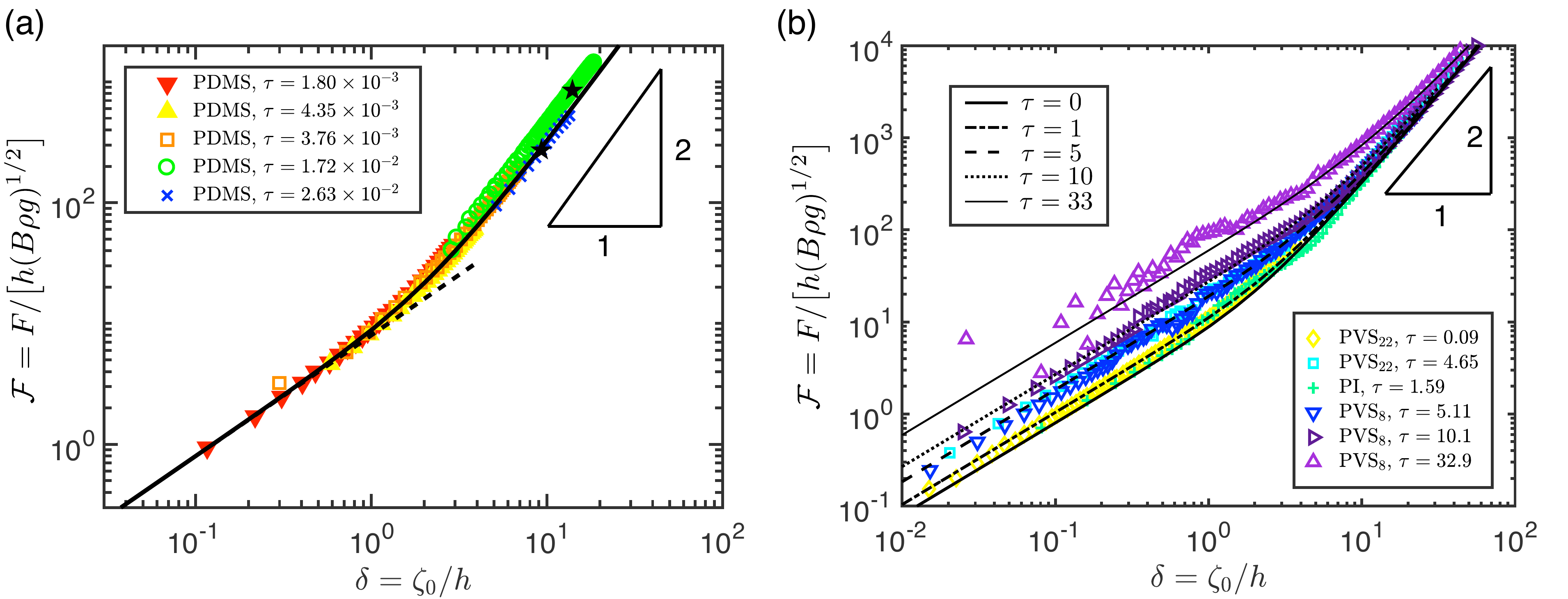}
\caption{(Color online). Localized dimensionless force, ${\cal F}=F/[h(B\rho g)^{1/2}]$, required to impose a given central deformation, $\delta=\zeta_0/h$, of a floating sheet  for a variety of mechanical bendabilities,  $\tau$. (a) Experimental results, obtained with $\tau \lesssim 10^{-2}$, are represented by markers (see table \ref{table} and legend for details) together with numerical results for the case $\tau=0$ (solid black curve) and the small loads result $\cF=8\wo$ (dashed line). For large loads, the  scaling predicted in \eqref{eqn:largedispforcelaw} is observed. The onset of wrinkling, which was observed for sheets of thickness $1.5$ and $2$\,mm, is indicated by a solid star. (b) Experimental results for $10^{-1} \lesssim \tau \lesssim 30$ are represented by markers (see table \ref{table} and legend for details) together with numerical results for $\tau=0, 1, 5, 10$ and $33$ (as indicated in the legend). The applied force in the experiments on PDMS were measured with an accuracy of $0.001$\,N and the indentation depth in experiments on PVS, PI and PC were measured with an accuracy of $0.1\,\mu$m, respectively. (Error bars are not shown on the plots for visual clarity.) } \label{fig:LoadDisp} 
\end{figure}

The system of equations \eqref{eqn:fvk1ndaxis}--\eqref{eqn:bc2axis} can readily be solved numerically using, for example, the MATLAB routine \texttt{bvp4c}. This numerical solution is computed on a finite domain, $[0,D/(2\lg)]$, where we use  $D/\lg = 2000$ to ensure that the domain is large enough that its finite size is not apparent when comparing with our analytical results (which are calculated with $D/\lg=\infty$). This yields predictions for the axisymmetric shape $Z(R)$ and the stresses within the sheet, and may also be used to determine the indentation force $\cF$ required to produce a given indentation depth $\wo$, since the first integral of \eqref{eqn:fvk1ndaxis} gives
\beq
\cF=-2\pi\lim_{R\to0}\left\{R\frac{\upd}{\upd R}\left[\frac{1}{R}\frac{\upd}{\upd R}\left(R\frac{\upd Z}{\upd R}\right)\right]\right\}.
\label{eqn:Fndaxis}
\eeq The force--displacement relationship calculated for $\tau=0$ is shown as the solid curve in figure~\ref{fig:LoadDisp}(a) along with the results from experiments, obtained for $\tau \lesssim 10^{-2}$ and detailed in \S\ref{sec:expt}(a). In figure~\ref{fig:LoadDisp}(b), the numerically determined force--displacement relationship is shown for a variety of values of $\tau$ together with experimental results for $10^{-1} \lesssim \tau \lesssim 30$, detailed in \S\ref{sec:expt}(b). Both the numerical and experimental results reveal the existence of two apparent regimes in the force--displacement law: for `small' displacements, $\delta \lesssim1 $, the displacement of the sheet scales linearly with the applied force, while for `large' displacements,  $\delta \gtrsim 1 $, the force scales with the square of the imposed displacement. Moreover, the results shown in figure~\ref{fig:LoadDisp}(b) also expose a dependence of the force--displacement on the mechanical bendability $\tau$ that appears to only be present for `small' displacements.
We therefore turn to try and understand these relationships analytically and to quantify more precisely what is meant by `small' and `large' displacements.

\definecolor{green1}{RGB}{255,255,0}
\definecolor{green2}{RGB}{127,255,0}
\definecolor{green3}{RGB}{0,255,0}
\definecolor{green4}{RGB}{0,255,127}
\definecolor{green5}{RGB}{0,255,255}
\definecolor{pink1}{RGB}{0,127,255}
\definecolor{pink2}{RGB}{0,0,255}
\definecolor{pink3}{RGB}{75,0,130}
\definecolor{pink4}{RGB}{148,0,211}
\begin{table}
\begin{center}
\begin{tabular}{ |c|c|c|c|c|c|c|}
\hline
Material & $h$ (mm) & $D$ (mm) &  $B$ ($\mathrm{Pa~m}^{3}$) & $\lg$ (mm) & $\tau$ & marker \\
\hline
PDMS & 9.0$\pm$0.02 & 931$\pm$0.5 & 1.67x10$^{-1}$ & 64.3 & 1.80x10$^{-3}$ & \textcolor{red}{$\blacktriangledown$} \\
PDMS & 5.5$\pm$0.02 & 890$\pm$0.5 & 3.81x10$^{-2}$ & 44.4 & 3.76x10$^{-3}$ & {\color{orange} $\square$} \\
PDMS & 5.0$\pm$0.02 & 480$\pm$0.5 & 2.86x10$^{-2}$ & 41.3 & 4.35x10$^{-3}$ & \textcolor{yellow}{$\blacktriangle$} \\
PDMS & 2.0$\pm$0.02 & 890$\pm$0.5 & 1.83x10$^{-3}$ & 20.8 & 1.72x10$^{-2}$ & \textcolor{green}{$\bigcirc$} \\
PDMS & 1.5$\pm$0.02 & 480$\pm$0.5 & 7.73x10$^{-4}$ & 16.8 & 2.63x10$^{-2}$ & \textcolor{blue}{$\times$} \\
PVS$_{22}$ & (929$\pm$8)$\times 10^{-3}$ & 89$\pm$0.5 & 7.14x10$^{-5}$ & 9.24 & 8.61x10$^{-2}$ & \textcolor{green1}{$\Diamond$} \\
PVS$_{22}$ & (367$\pm$7)$\times 10^{-3}$ & 89$\pm$0.5 & 4.39x10$^{-6}$ & 4.60 & 3.47x10$^{-1}$ & \textcolor{green2}{$\times$} \\
PVS$_{22}$ & (169$\pm$5)$\times 10^{-3}$ & 89$\pm$0.5 & 4.30x10$^{-7}$ & 2.57 & 1.11 & \textcolor{green3}{$\ast$} \\
PVS$_{22}$ & (100$\pm$1)$\times 10^{-3}$ & 89$\pm$0.5 & 8.90x10$^{-8}$ & 1.74 & 2.44 & \textcolor{green4}{$\bigcirc$} \\
PVS$_{22}$ & (65$\pm$1)$\times 10^{-3}$ & 89$\pm$0.5 & 2.44x10$^{-8}$ & 1.26 & 4.65 & \textcolor{green5}{$\square$} \\
PVS$_8$ & (244$\pm$13)$\times 10^{-3}$ & 89$\pm$0.5 & 3.23x10$^{-7}$ & 2.40 & 1.14 & \textcolor{pink1}{$\lhd$} \\
PVS$_8$ & (90$\pm$3)$\times 10^{-3}$ & 89$\pm$0.5 & 1.62x10$^{-8}$ & 1.13 & 5.11 & \textcolor{pink2}{$\triangledown$} \\
PVS$_8$ & (57$\pm$4)$\times 10^{-3}$ & 89$\pm$0.5 & 4.12x10$^{-9}$ & 0.805 & 10.1 & \textcolor{pink3}{$\rhd$} \\
PVS$_8$ & (26$\pm$4)$\times 10^{-3}$ & 89$\pm$0.5 & 3.91x10$^{-10}$ & 0.447 & 32.9 & \textcolor{pink4}{$\triangle$} \\
PI & (8.5$\pm2$)$ \times 10^{-3}$ & 200$\pm$0.5 & 2.14x10$^{-7}$ & 2.16 & 1.59 & \textcolor{green4}{$+$} \\
\hline
\end{tabular}
\caption[Elastic sheet properties]{The properties of the elastic sheets used in the investigation of the axisymmetric deformation of a floating sheet subject to a localized load, including: the material,  sheet thickness ($h$), diameter ($D$), and bending stiffnesses ($B=Eh^{3}/[12(1-\nu^{2})]$), together with the calculated values of the \emph{elasto-gravitational} length ($\lg$) and mechanical bendability, $\tau$, defined in \eqref{eqn:tauDefn}. The table also includes the data marker used to denote experimental results with each elastic sheet in figures~\ref{fig:LoadDisp} and \ref{fig:stiffness}.}
\label{table}
\end{center}
\end{table}

\subsection{Small displacement theory}

For small displacements, we expect that the stress within the sheet should remain axisymmetric (independent of $\theta$) being close to the unperturbed value set by the far-field tension (or mechanical bendability), and, further, that the vertical displacement will be everywhere small. Following a similar analysis for the indentation of pressurized elastic shells \cite{Vella2012}, we seek small deviations from this uniform tension state and so let $\Psi=\tau R^2/2+\dPsi$ and $Z=\dW$.  Hence \eqref{eqn:fvk1ndaxis} and \eqref{eqn:fvk2ndaxis} may be linearized to give 
\begin{equation}
\nabla^4\dW-\tau \nabla^2\dW+ \dW=-\frac{\cF}{2\pi}\frac{\delta(R)}{R}
\label{fvk1:smalldef}
\end{equation} and 
\begin{equation}
R\frac{\upd }{\upd R}\left[\frac{1}{R}\frac{\upd }{\upd R}\left(R\frac{\upd\dPsi}{\upd R}\right)\right]=0.
\label{fvk2:smalldef}
\end{equation} 

The solution of \eqref{fvk1:smalldef} can be found by noting that solutions of the Helmholtz equation $\nabla^2\dW=\lambda \dW$ are also solutions of \eqref{fvk1:smalldef} if
\beq
\lambda^2-\tau\lambda+1=0,
\eeq and hence that
\beq
\lambda=\lambda_\pm=\tfrac{1}{2}\left(\tau\pm\sqrt{\tau^2-4}\right).
\eeq We therefore have that the relevant general solution for small vertical deflections is
\beq
\dW=\alpha K_0(\lp^{1/2}R)+\beta K_0(\lm^{1/2}R),
\eeq where $K_0(x)$ is the modified Bessel function of zeroth order and the constants $\alpha$ and $\beta$ need to be chosen to satisfy the boundary conditions as $R\to 0$. (The conditions as $R\to \infty$ have  already been satisfied by our choice of the solution of the Helmholtz equation --- we have neglected the possibility of any solutions $\propto I_0(x)$, which would diverge as $x\to\infty$.) We find that
\beq
\dW=-\frac{2\wo}{\log(\lm/\lp)}\left[K_0(\lp^{1/2}R)-K_0(\lm^{1/2}R)\right].
\label{eqn:smallDprof}
\eeq 
It is possible to repeat this calculation to account for the effect of a finite-sized indenter, $r_{\mathrm{in}}$. We find that the prefactor in \eqref{eqn:smallDprof} is correct to $O(r_{\mathrm{in}}/\lg)$ and so the limit of a point indenter, $r_{\mathrm{in}}/\lg\to0$, is regular. (This regularity is a result of the finite bending stiffness of the sheet, and is distinct from the indentation of a membrane, where a logarithmic dependence on the indenter radius was found \cite{Vella2017} analytically, although a power-law correction has also been claimed \cite{Bernal2011}.) In the majority of our experiments, $r_{\mathrm{in}}/\lg\lesssim0.1\ll1$ and so the effect of indenter size may be neglected. Finally, we note  that, at this order, the perturbed Airy stress function $\dPsi=0$: from \eqref{fvk2:smalldef}, $\dPsi=AR^2+B$, which cannot satisfy the boundary conditions unless $A=B=0$.

To compute the force required to produce the displacement in \eqref{eqn:smallDprof} we use \eqref{eqn:Fndaxis}, which gives that
\begin{equation}
\cF=K_1\wo
\label{eqn:Fk1}
\end{equation} where
\begin{equation}
K_1=2\pi\frac{(\tau^2-4)^{1/2}}{\atanh\bigl[\bigl(1-4/\tau^2\bigr)^{1/2}\bigr]}
\label{k1:full}
\end{equation} is the dimensionless `stiffness', or spring constant, of the floating sheet.  

\begin{figure}
	\centering
		\includegraphics[width=0.7\textwidth]{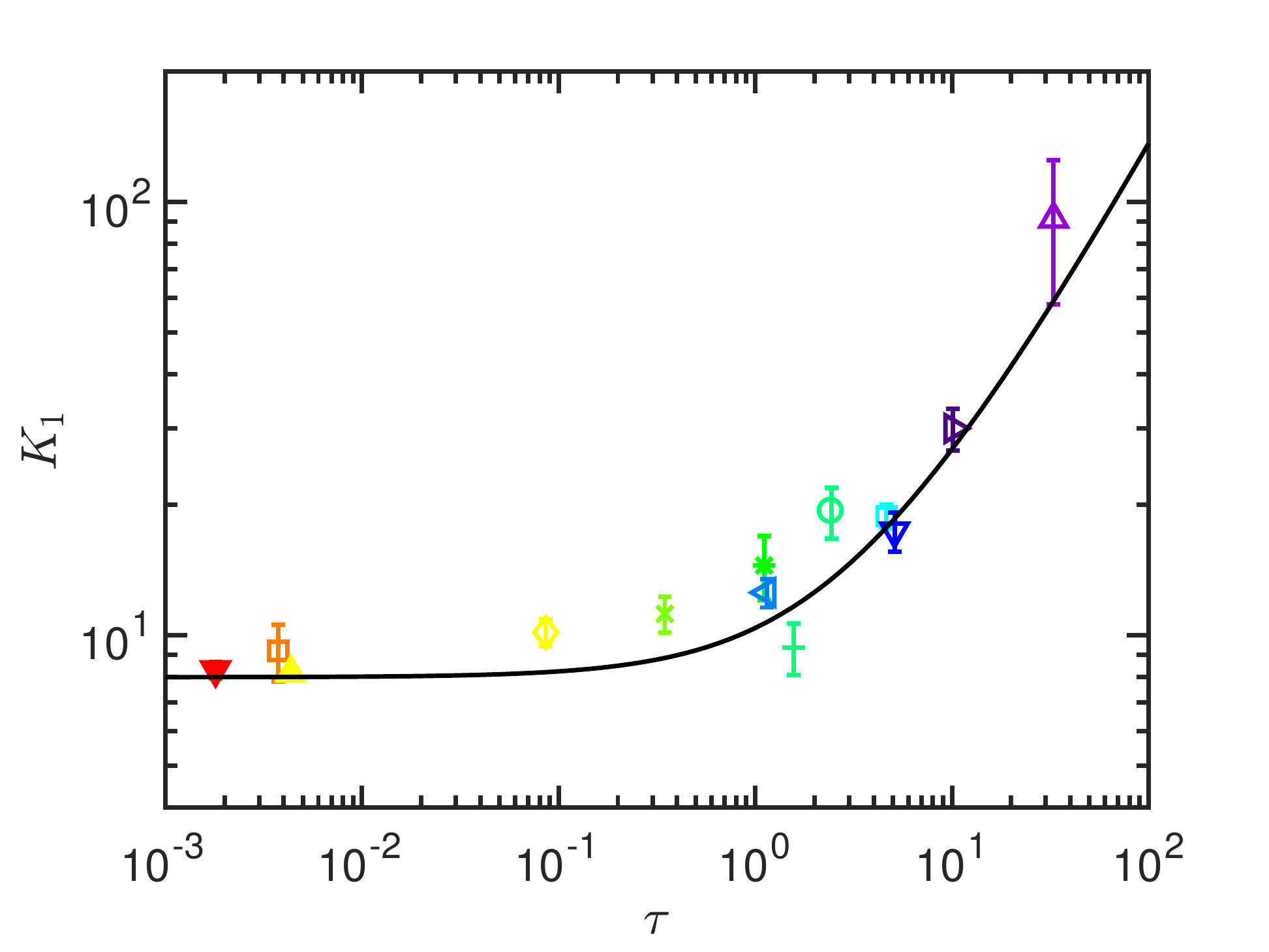}
	\caption{The small indentation, $\delta<1$ spring stiffness $K_{1}$ measured as a function of the mechanical bendability $\tau$. The data markers correspond to the force--displacement  experiments detailed in Table \ref{table} and the black curve corresponds to the theoretical prediction \eqref{k1:full}. Error bars represent the standard deviation of the measured values of $K_{1}$ calculated from force-displacement measurements for $\delta<1$.}
	\label{fig:stiffness}
\end{figure}

It is important to note that the dimensionless stiffness \eqref{k1:full} is  a function solely of  $\tau$, as defined in \eqref{eqn:tauDefn}.  In figure~\ref{fig:stiffness}, experimental values of $K_{1}$ are shown as a function of $\tau$ alongside the theoretical result \eqref{k1:full}.  (For definiteness, experimental calculations of the spring constant used experimental data with $\delta<1$.)

We note that in the limits of small and large mechanical bendabilities the dimensionless stiffness takes the values
\begin{equation}
K_1\sim\begin{cases}
8,\quad\tau\ll1\\
2\pi\frac{\tau}{\log (4\tau)},\quad\tau\gg1.
\end{cases}
\end{equation}  For  $\tau\ll1$ the stiffness of the sheet becomes insensitive to the value of that tension (since the restoring force is provided predominantly by the bending stiffness of the sheet). In dimensional terms, we have that for $\tau\ll1$
\beq
F\approx 8 \frac{B}{\lg^2}\zeta_0=8 (B\rho g)^{1/2}\zeta_0,
\eeq which is a result derived first by Hertz \cite{Hertz1884}. For large mechanical bendability, $\tau\gg1$, however, the stiffness of the sheet is instead dominated by the surface tension of the interface; the corresponding dimensional result is
\beq
F\approx\frac{2\pi\glv}{\log(4\tau)} \zeta_0.
\eeq In this limit, the bending stiffness of the sheet  enters only via a logarithmic correction. 

\subsection{Large displacement scaling analysis \label{subsec:LDT}}

The theoretical analysis just presented for small displacements relied on the stress state within the sheet remaining close to its pre-indentation levels. However, as the indentation, $\zeta_0$, increases the sheet is forced to stretch over a horizontal region of size $\lcurv$ (with $\lcurv$ currently unknown). This stretching induces a strain $\sim(\zeta_0/\lcurv)^2$ and hence costs an elastic energy $\sim Eh (\zeta_0/\lcurv)^4\lcurv^2= Eh\zeta_0^4/\lcurv^2$. Since this elastic energy decreases as the sheet stretches over a greater horizontal distance (the strain is smaller), it is tempting to assume that $\lcurv=R_{\mathrm{sheet}}$, the radius of the sheet. However, such a deformation is extremely expensive in terms of the gravitational potential energy of the liquid that is displaced, $\sim \rho g \zeta_0^2\lcurv^2$. Instead, an indentation-dependent horizontal scale $\lcurv\sim (Eh\zeta_0^2/\rho g)^{1/4}$ emerges that minimizes the sum of gravitational and elastic energies. 
Using this estimate of  $\lcurv$ in the above energy estimates, we find that the total energy of the system then scales like $(Eh\rho g)^{1/2}\zeta_0^3$, which must balance the work done in indentation, $F\zeta_0$. This argument predicts that the indentation force $F\sim(Eh \rho g)^{1/2}\zeta_0^2$, or in dimensionless terms that
\beq
\cF\sim \wo^2
\label{eqn:largedispforcelaw}
\eeq with a pre-factor that is independent of $\tau$. This scaling is  consistent with both numerical and experimental results shown in figure~\ref{fig:LoadDisp}. However, it seems possible that this scaling might fail with $\tau\gg1$; for example, the relevant stretching energy might be that in the flat portion of the membrane, $r\gtrsim\lcurv$, rather than that induced by the out of plane deformation. We shall carefully consider this possibility  shortly.


\begin{figure}
\centering
\includegraphics[width=\textwidth]{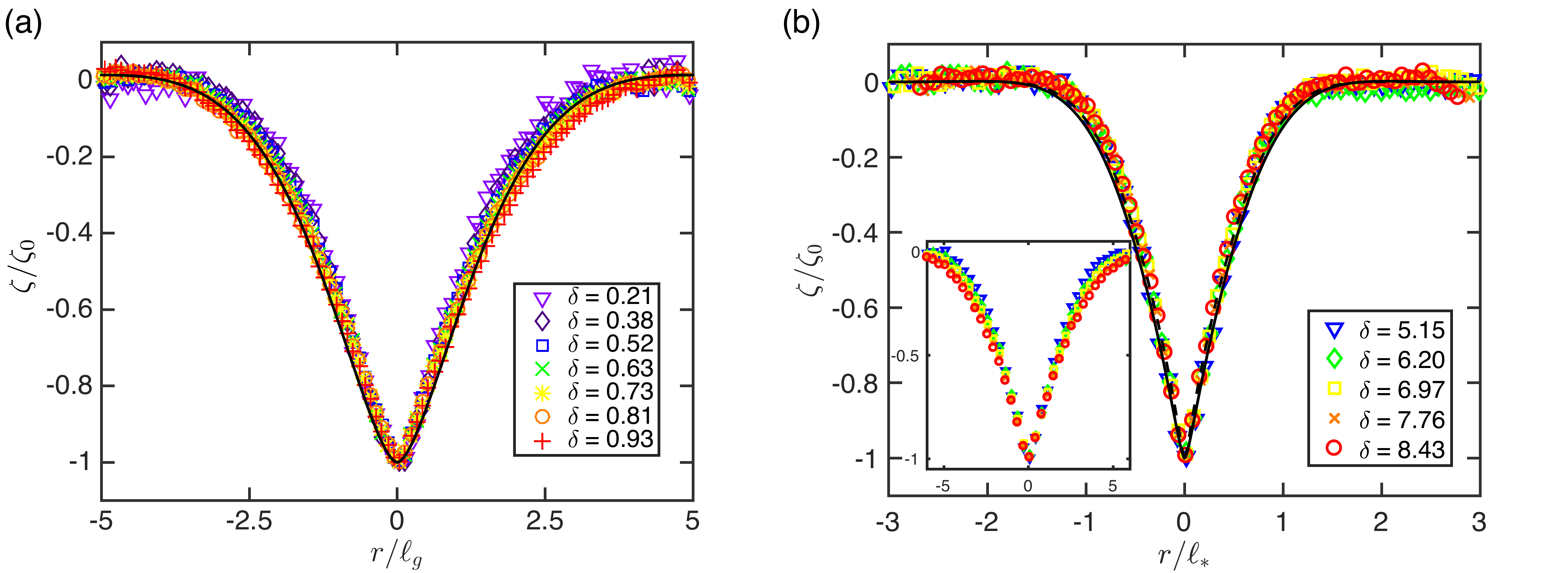}
\caption[Deflection profiles]{(Color online). Deformation profile of a floating elastic sheet with low mechanical bendability, $\tau\ll1$, and subject to a localized load for various indentation depths (as indicated in each legend). (a) For small displacements, $\delta=\zeta_{0}/h <1$, the normalized vertical displacement $\zeta/\zeta_0$ is plotted as a function of the radial distance scaled by the elasto-gravity bending length $\lg$. The analytical prediction, \eqref{eqn:smallDprof}, is also shown for the case $\tau=0$ (solid curve). (b) For large indentation depths, $\delta=\zeta(0)/h>1$, the normalized vertical  displacement $\zeta/\zeta_0$ is plotted as a function of the radial distance  scaled by the emergent horizontal scale, $\ell_\ast = (Eh\zeta_{0}^{2}/\rho g)^{1/4}=\lg\delta^{1/2}[12(1-\nu^2)]^{1/4}$; also shown are the numerically obtained predictions for $\delta=5$ (dashed curve) and $\delta=8$ (solid curve). The inset shows how these data would collapse if $\lg$ were used to rescale horizontal lengths, as is appropriate for small indentation depths. The experimental data was obtained for PDMS sheets with (a)  $h = 5$\,mm and  (b)  $h=1.5$\,mm; however, for all of the data presented, the deformation of the sheet remains axisymmetric so that imaging the deflection of a line drawn across the centre of the sheet provides a measure of the deformation independent of azimuthal angle.
}
\label{fig:DeflectionProfiles}
\end{figure}

\begin{figure}
\centering
\includegraphics[width=0.95\textwidth]{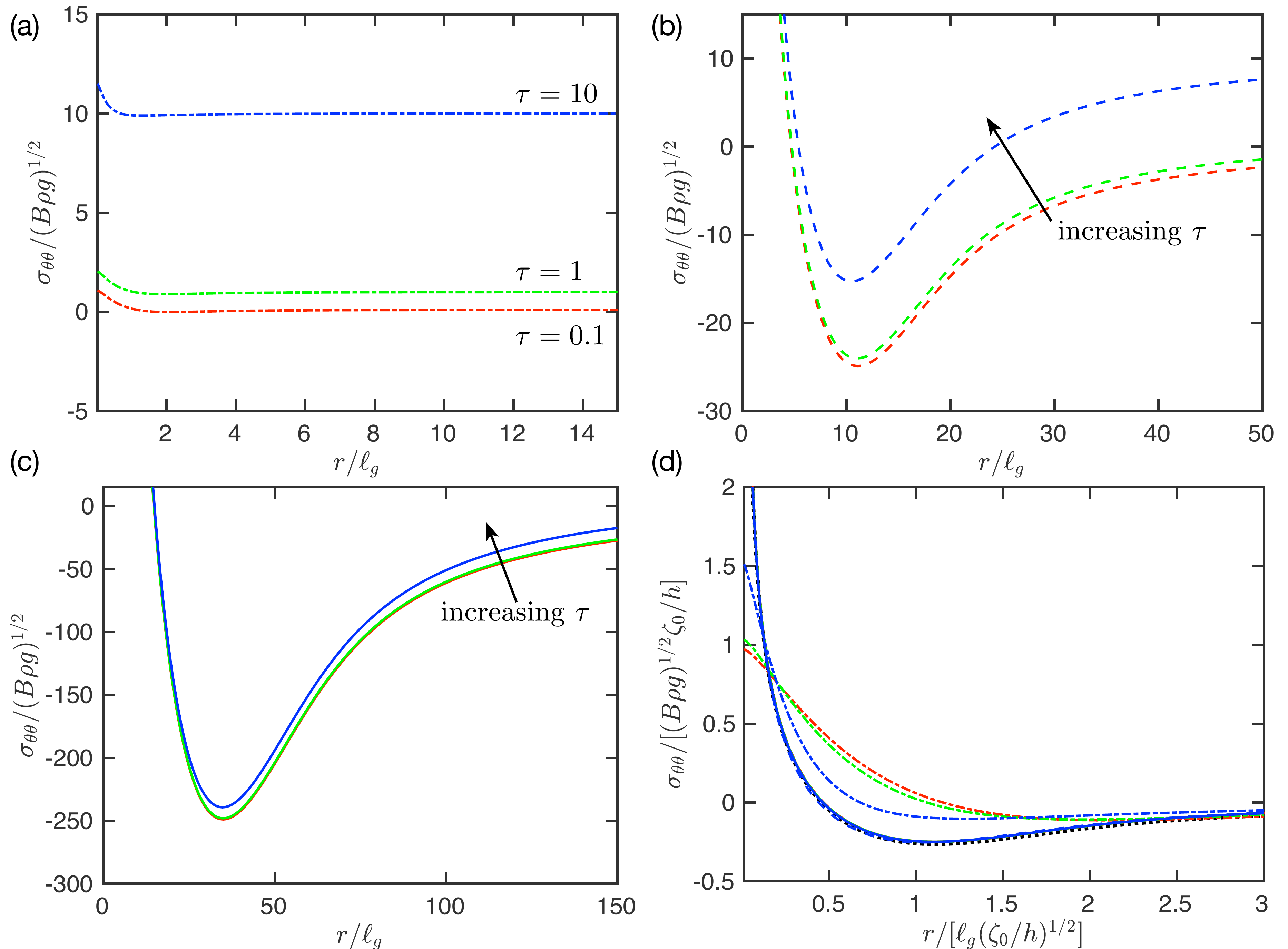}
\caption[Stress profiles]{(Color online). Numerically determined  profiles of the hoop stress in the sheet for different indentation depths, $\wo$, and mechanical bendabilities, $\tau$. Results are shown for $\tau=0.1$ (red curves), $\tau=1$ (green curves) and $\tau=10$ (blue curves). (a) At relatively small indentation depths (here $\delta=1$) the stress is approximately uniform and close to the initial stress, $\tau$, except very close to the indenter. For larger indentation depths the region in which the stress changes significantly grows and $\sqq<0$ in an annular region that also grows: in (b) $\delta=10^2$ and in (c) $\delta=10^3$. In both (b) and (c) curves are shown for $\tau=0.1$, $1$ and $10$  with the direction of increasing $\tau$ indicated by arrows.  (d) Rescaling the results in (a)-(c) as suggested by the scaling analysis, and in particular \eqref{eqn:rescale}, shows that the hoop stress profile approaches a universal profile (black dotted curve) when $\wo\gg1$ and $\wo/\tau\gg1$.
}
\label{fig:StressProfiles}
\end{figure}

Profiles of the deflection of the PDMS sheets measured as a function of radial position are shown in figure~\ref{fig:DeflectionProfiles} for various ratios of indentation displacement to sheet thickness, $\delta = \zeta_{0}/h$. Naturally, an increase in the applied load resulted in greater deformation of the elastic sheet. For the results shown in figure~\ref{fig:DeflectionProfiles}(a) the indentation of the sheet is less than the thickness of the sheet, $\delta<1$, while in figure~\ref{fig:DeflectionProfiles}(b) the indentation is greater than the sheet thickness, $\delta>1$ (though in all cases plotted in figure \ref{fig:DeflectionProfiles}, the measured deformation of the sheet remains axisymmetric). For  small deformations, the normalized indentation displacement, $\zeta/\zeta_0$, is plotted as a function of the radial distance scaled by the elasto-gravity bending length, $r/\lg$,  in figure \ref{fig:DeflectionProfiles}(a); we observe good collapse, particularly with the smallest $\delta \ll 1$. In contrast, for large deformations, shown in \ref{fig:DeflectionProfiles}(b), the normalized indentation displacement $\zeta/\zeta_0$ is plotted as a function of the radial distance $r$ scaled by the horizontal scale that emerges from the large displacement scaling analysis  $\ell_\ast \sim (Eh\zeta_{0}^{2}/\rho g)^{1/4}$ --- we see that plotting the data in this way provides a better collapse in this regime than would be obtained by using $r/\lg$ (see inset of figure~\ref{fig:DeflectionProfiles}(b)). For both small and large displacements, we see that the vertical deflection of the edge of the sheet is negligible, though this is expected to break down for very large indentations, i.e.~once  $\ell_\ast\sim R_{\mathrm{sheet}}$.

Figure \ref{fig:StressProfiles}(a)-(c) presents numerically determined profiles of the hoop stress, $\sqq$, within the sheet for $\tau=0.1, 1$ and $10$, and different values of the indentation depth $\wo$. These stress profiles show that the stress differs from the applied (interfacial) tension only near the indenter, but also that the indenter's region of influence grows with increasing indentation depth. The key feature of these plots is, however, that the hoop stress becomes increasingly compressive as $\wo$ increases --- this compression ultimately leads to wrinkling, as we shall see in \S\ref{sec:wrinkling}.

The scaling law that led to \eqref{eqn:largedispforcelaw} predicts that the stresses within the sheet, $\sigma_{ij}\sim\wo$. It is therefore natural to test this scaling law by rescaling the numerical results of figure \ref{fig:StressProfiles}(a)-(c) using a lateral length scale $\wo^{1/2}$ and stress scale $\wo$. This collapse is shown in figure \ref{fig:StressProfiles}(d) and supports the validity of the scaling argument as already presented. However, two dimensionless groups, both $\tau$ and $\wo$, influence the behaviour and so it seems plausible that some $\tau$-dependence may remain. To test this possibility, we use the rescalings suggested by our scaling analysis to rescale the full, axisymmetric dimensionless problem \eqref{eqn:fvk1ndaxis}--\eqref{eqn:fvk2ndaxis}. In particular, we let
\beq
\tR=R/\wo^{1/2},\quad\tZ=Z(R)/\wo,\quad \tPsi=\left(\Psi-\tfrac{1}{2}\tau R^2\right)/\wo^2
\label{eqn:rescale}
\eeq and find that the system of equations \eqref{eqn:fvk1ndaxis}--\eqref{eqn:bc2axis} becomes
\beq
\wo^{-2}\frac{1}{\tR}\frac{\upd}{\upd \tR}\left\{\tR\frac{\upd}{\upd \tR}\left[\frac{1}{\tR}\frac{\upd}{\upd \tR}\left(\tR\frac{\upd \tZ}{\upd \tR}\right)\right]\right\}-\frac{1}{\tR}\frac{\upd}{\upd \tR}\left[\frac{\upd \tZ}{\upd \tR}\left(\frac{\tau}{\wo}\tR+\frac{\upd \Psi}{\upd R}\right)\right]=-\tZ-\frac{\cF}{2\pi\wo^2}\frac{\delta(\tR)}{\tR}
\label{eqn:fvk1ndaxisRS}
\eeq and
\beq
\tR\frac{\upd}{\upd \tR}\left\{\frac{1}{\tR}\frac{\upd}{\upd \tR}\left[\tR\left(\frac{\tau}{\wo}\tR+\frac{\upd\tPsi}{\upd \tR}\right)\right]\right\}=-6(1-\nu^2)\left(\frac{\upd \tZ}{\upd \tR}\right)^2,
\label{eqn:fvk2ndaxisRS}
\eeq subject to
\beq
\tZ(0)=-1,\quad \tZ'(0)=\lim_{\tR\to0}\bigl[\tR\tPsi''-\nu \tPsi' \bigr]=\tPsi(0)=0,
\label{eqn:bc1axisRS}
\eeq and far-field conditions
\beq
\tZ,\tZ'\to0,\quad\tPsi\to0 \quad\quad (\tR\to\infty).
\label{eqn:bc2axisRS}
\eeq

This rescaling shows that the effect of the bending stiffness (the first term on the LHS of \eqref{eqn:fvk1ndaxisRS}) may be neglected for $\wo\gg1$, apart from a small boundary layer near the origin. Interestingly, this rescaling also reveals that the effect of the mechanical bendability is perturbative, provided that the indentation depth is sufficiently large that $\tau/\wo\ll1$. Therefore in the limit $\wo\gg\max\{\tau,1\}$, a universal problem is recovered and the problem, including the force law \eqref{eqn:largedispforcelaw}, is indeed independent of the mechanical bendability $\tau$, even for $\tau\gg1$.

\subsection{Transition from small to large displacements}
\label{subsec:transition}

To understand what is meant by large and small displacements, we now compare the two force laws given by \eqref{eqn:Fk1} and \eqref{eqn:largedispforcelaw}. We expect that these two forces become of the same order when
\beq
\wo\sim K_1(\tau),
\label{eqn:deltaSwitch}
\eeq and hence that the transition between the linear and quadratic regimes will occur when the dimensionless displacement is comparable to the dimensionless stiffness of the sheet, i.e.~$\wo=O[K_1(\tau)]$. Since $K_1(\tau)\sim \max\{1,\tau/\log(\tau)\}$, we note that the criterion \eqref{eqn:deltaSwitch} for the transition between small and large indentation depths is also consistent with the condition $\wo\gg\max\{\tau,1\}$ for which the rescaled problem \eqref{eqn:fvk1ndaxisRS}--\eqref{eqn:bc2axisRS} becomes universal.

The  analysis of this section has characterized the axisymmetric behaviour of the sheet. As the load, $\cF$, or, equivalently, the indentation depth $\wo$, increase, the stresses within the sheet are changed from the uniform, homogeneous tension applied by surface tension initially. As might be expected, the application of a load generally stretches the sheet, increasing the magnitude of the stress. However, indentation also acts to pull material within the sheet to a a smaller radial coordinate: to fit within this smaller circle, the hoop stress, $\sqq$, becomes relatively compressive (see for example figure \ref{fig:StressProfiles} and  stress profiles for the membrane, $\tau=\infty$, case \cite{Vella2015}). The degree of relative compression increases with indentation depth and, at sufficiently large $\wo$, the hoop stress becomes absolutely compressive, $\sqq<0$. Very thin membranes ($\tau\gg1$) offer very little resistance to bending \cite{Vella2015}, and so this compression signals the onset of wrinkling. However, to determine this onset of wrinkling for finite mechanical bendability requires more detailed consideration, and it is to this that we now turn.

\section{Large-amplitude deformation: the onset of wrinkling\label{sec:wrinkling}}

In the  limit of infinite mechanical bendability, $\tau=\infty$, it was shown previously \cite{Vella2015} that the hoop stress first becomes compressive when
\beq
\zeta_0=\zeta_0^{(c)}\approx11.75\lc\left(\frac{\glv}{Eh}\right)^{1/2},
\eeq which in the non-dimensionalization used here reads
\beq
\woc=\frac{\zeta_0^{(c)}}{h}\approx \frac{11.75}{[12(1-\nu^2)]^{1/2}}\tau.
\label{eqn:largetauOnset}
\eeq It was also shown experimentally that very thin membranes do indeed wrinkle at this indentation depth, to within experimental resolution.

However, sheets of finite mechanical bendability are of most interest here and do, by definition, have a finite resistance to bending. As such, they may accommodate some compressive stresses before buckling. We therefore expect that there will be a $\tau$-dependent critical indentation depth $\woc(\tau)$ at which a wrinkled solution first exists. Determining this critical indentation depth should also reveal the properties of the wrinkle pattern at onset. We  therefore focus on the behaviour of the system close to the threshold of wrinkling: this is a `Near Threshold' analysis, rather than the `Far from Threshold' analysis in which the stress field is fundamentally changed to relax compression \cite{Davidovitch2011,Vella2015}.

\subsection{Linear stability analysis}

We seek a solution of \eqref{eqn:fvk1nd} and \eqref{eqn:fvk2nd} that is a small perturbation of the axisymmetric solution found in \S\ref{sec:axisymm}. We therefore let
\begin{eqnarray}
Z(R,\theta)&=&\Wnought(R)+ \Wone(R)\cos n\theta+...,\nonumber\\
\chi(R,\theta)&=&\Chinought(R)+ \Chione(R)\cos n\theta+... .\nonumber
\end{eqnarray} Substituting this ansatz into \eqref{eqn:fvk1nd} and \eqref{eqn:fvk2nd}, we find that at leading order we recover the axisymmetric membrane problem considered in \S\ref{sec:axisymm}, which takes the form of \eqref{eqn:fvk1ndaxis}--\eqref{eqn:fvk2ndaxis} with $\Psi\to\Chinought$ and $Z\to\Wnought$, i.e.
\begin{eqnarray}
\nabla^4\Wnought&=&\frac{1}{R}\frac{\upd }{\upd R}\left(\frac{\upd \Wnought}{\upd R}\frac{\upd \Chinought}{\upd R}\right)-\Wnought-\frac{\cF}{2\pi}\frac{\delta(R)}{R},\nonumber\\
\nabla^4\Chinought&=&-\frac{6(1-\nu^2)}{R}\frac{\upd }{\upd R}\left(\frac{\upd \Wnought}{\upd R}\right)^2.\nonumber
\end{eqnarray} However, at next order, and after retaining only those terms that are linear in the perturbation,  we find that
\begin{eqnarray}
\frac{1}{R^2}\left[\Chinought''\left(R\Wone'-n^2\Wone\right)+R\Chinought'\Wone''+\Wnought''\left(R\Chione'-n^2\Chione\right) +R\Wnought'\Chione''\right] \nonumber \\
= \Wone + {\cal L}_n^2\left\{\Wone\right\}
\label{fvk1:ndim}
\end{eqnarray} and
\begin{equation}
{\cal L}_n^2\left\{\Chione\right\}+\frac{12(1-\nu^2)}{R^2}\left[R\Wnought'\Wone''+(R\Wone'-n^2\Wone)\Wnought''\right]=0,
\label{fvk2:ndim}
\end{equation} where $f'$ denotes differentiation with respect to $R$ and the operator ${\cal L}_n$ is defined by
\beq
{\cal L}_n\left\{f\right\}=\left(\frac{\upd^2}{\upd R^2}+\frac{1}{R}\frac{\upd}{\upd R}-\frac{n^2}{R^2}\right)f.
\label{eqn:linstab1}
\eeq For practical purposes, it is useful to note that
\beq
{\cal L}_n^2\left\{f\right\}=f''''(R)+\frac{2}{R}f'''(R)-\frac{2n^2+1}{R^2}f''(R)+\frac{2n^2+1}{R^3}f'(R)+\frac{n^2(n^2-4)}{R^4}f(R).
\label{eqn:linstab2}
\eeq

Equations \eqref{fvk1:ndim} and \eqref{fvk2:ndim} are to be solved subject to the boundary conditions that the perturbation to the displacement and its slope must vanish both at the indenter and as $R\to \infty$, that is
\beq
\Wone(0)=\Wone'(0)=0, \quad \Wone(R\to\infty) = \Wone'(R\to\infty) = 0.
\label{eqn:ZetaBCs}
\eeq Considering the condition that the components of the displacement have to vanish at the indenter gives
\beq
\lim_{R\to0}\left[R\Chione''-\nu\Chione'+\nu n^2\frac{\Chione}{R}\right]=0,
\label{eqn:ChiBCs}
\eeq as well as $\Chione(0)=0$.

The problem \eqref{fvk1:ndim}--\eqref{eqn:ChiBCs} is a quadratic eigenvalue problem for $n^2$ \cite{Tisseur2001}. We solve this problem numerically by first solving the axisymmetric problem via relaxation (using \texttt{bvp4c} in MATLAB). With this solution, we then discretize the resulting linear equations for $\Chione$ and $\Wone$ using centred finite differences and solve the resulting quadratic eigenvalue problem by restricting $n$ to be an integer and determining  the smallest value of $\wo$ for which the linear system has vanishing determinant. This gives a range of critical indentation depths $\woc(n)$, which can then be minimized to give the smallest value of $\wo$ at which wrinkles may occur, as well as the corresponding number of wrinkles at the onset of wrinkling, $\nonset$.  The results of this numerical  analysis, together with the results from experiments focussed on the onset and form of wrinkling, are detailed below.

\subsection{Experimental investigations of the onset of wrinkling}

\begin{figure}
	\centering
		\includegraphics[width = 0.8\textwidth]{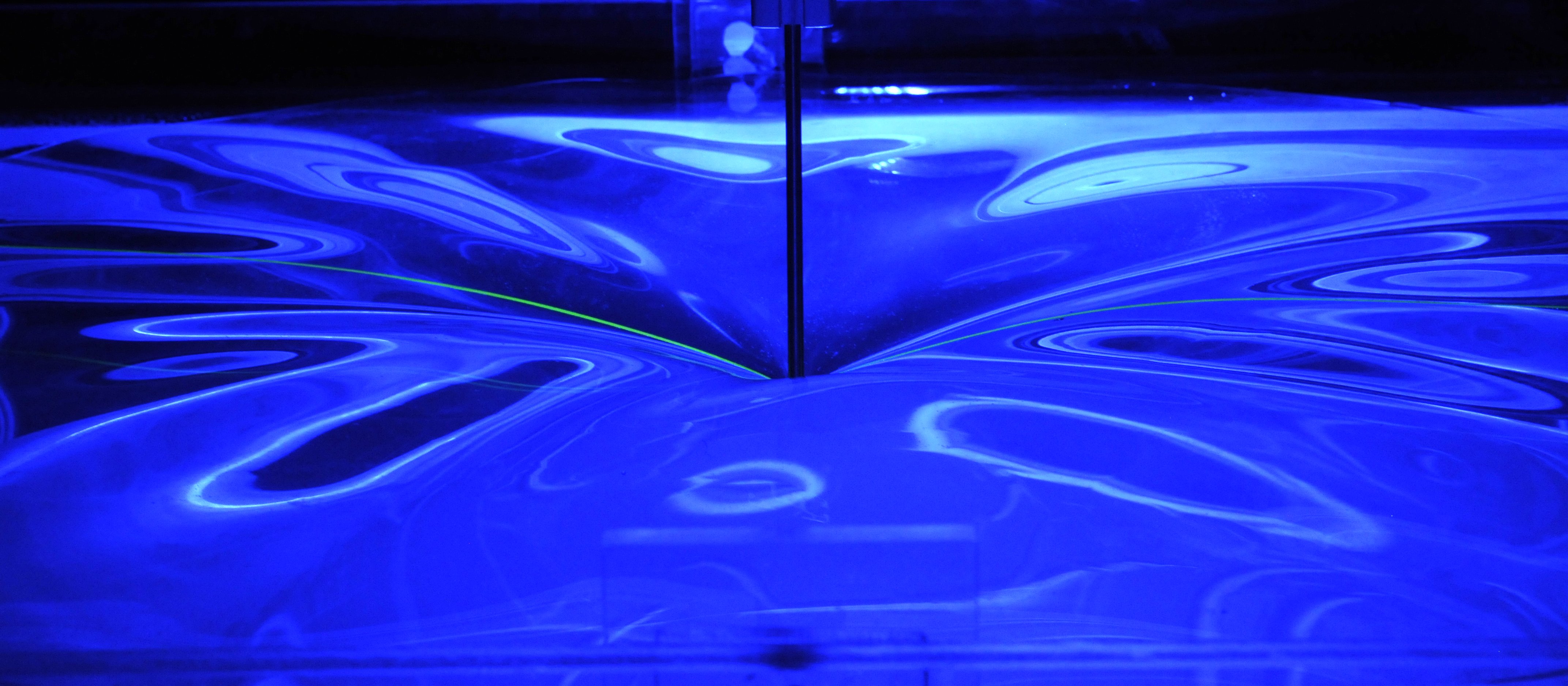}
	\caption[Wrinkle]{(Color online). Image of the wrinkles that result from the indentation of a floating PDMS sheet for $\wo \gg \woc$. (Here $h=2$\,mm and $D=890$\,mm.)}
	\label{fig:Wrinkle}
\end{figure}

To explore the onset of wrinkling, further experiments were performed on the elastic sheets described in \S\ref{sec:expt}. An image of the relatively large amplitude wrinkles observed with a thick PDMS sheet, for $\wo \gg \woc$, is shown in figure~\ref{fig:Wrinkle}. The key parameters of interest here, however, are those determined by our Near Threshold analysis, namely the critical indentation depth at which wrinkling first occurs, $\delta_c$, and the number of wrinkles present at the onset of wrinkling, $\nonset$. (We do not investigate how the wrinkle patterns evolve beyond onset, which has been studied in detail for highly bendable sheets \cite{Paulsen2016}.) The physical properties of the sheets used in these experiments are summarized in Table \ref{table2}. 

As before,  relatively thick PDMS sheets and thin PVS sheets (material properties as detailed in Table \ref{table2}) were floated on water and indented until the wrinkles became visible so that the number of wrinkles could be counted by eye.  Additional experiments were performed on a Polyimide (PI) sheet (detailed in \S\ref{sec:expt}(b)) and a Polycarbonate (PC) sheet of $h = 2.2$\,$\mu$m and $E = 2.73$\,GPa and $\nu = 0.37$ (supplied by Goodfellow, Cambridge). The experiments performed on the PI and PC sheets were performed with the sheets floating on water ($\glv=72.81\pm0.12$\,dyn/cm) and also on a well-mixed solution of water and washing-up liquid ($\glv=24.85\pm0.12$\,dyn/cm, measured as described previously).  The variation in the sheet thickness and material, as well as of the liquid--vapour surface tension $\glv$ permitted the variation of $\tau$: here we report experiments with $10^{-2}\lesssim\tau\lesssim30$.

\begin{table}
\begin{center}
\begin{tabular}{ |c|c|c|c|c|c|c|c|}
\hline
Material & $h\mathrm{~(\mu m)}$ & $D$ (mm) &  $B \mathrm{~(Pa\,m^{3})}$ & $\lg$ (mm) & $\lc$ (mm) & $\tau$ & marker \\
\hline
PDMS & 2000 & 890$\pm$0.5 & $1.83\times10^{-3}$ & 20.8 & 2.72 & $1.72\times10^{-2}$ & \textcolor{red}{$\bullet$}  \\
PDMS & 1500 & 480$\pm$0.5 & $7.73\times10^{-4}$ & 16.8 & 2.72 & $2.63\times10^{-2}$ & \textcolor{red}{$\bullet$} \\
PI & 8.5$\pm2.0$  & 200$\pm$0.5 & $2.14\times10^{-7}$ & 2.16 & 2.72 & 1.59 & \textcolor{red}{$\bullet$} \\
PC &$2.2\pm0.6$ & 118$\pm$0.5 & $3.80\times10^{-9}$ & 0.79 & 2.72 & 11.9 & \textcolor{red}{$\bullet$} \\
PI & $8.5\pm2.0$ & 200$\pm$0.5 & $2.14\times10^{-7}$ & 2.16 & 1.59 & 5.43$\times 10^{-1}$ & \textcolor{blue}{$\blacktriangledown$} \\
PC & $2.2\pm0.6$ & 118$\pm$0.5 & $3.80\times10^{-9}$ & 0.79 & 1.59 & 4.07 & \textcolor{blue}{$\blacktriangledown$} \\
PVS$_8$ & $191\pm3$ & 89$\pm$0.5 & $1.56\times10^{-7}$ & 2.00 & 2.72 & 1.84 & \textcolor{green}{$\blacksquare$} \\
PVS$_8$ & $189\pm2$ &89$\pm$0.5 & $1.51\times 10^{-7}$ & 1.98 & 2.72 & 1.87 & \textcolor{green}{$\blacksquare$} \\
PVS$_8$ & $83\pm5$ & 89$\pm$0.5 & $1.28\times10^{-8}$ & 1.07 & 2.72 & 6.43 & \textcolor{green}{$\blacksquare$} \\
PVS$_8$ & $67\pm5$ & 89$\pm$0.5 & $6.72\times 10^{-9}$ & 0.91 & 2.72 & 8.84 & \textcolor{green}{$\blacksquare$} \\
PVS$_8$ & $48\pm5$ & 89$\pm$0.5 & $2.47\times10^{-9}$ & 0.71 & 2.72 & 14.6 & \textcolor{green}{$\blacksquare$} \\
PVS$_8$ & $48\pm3$ & 89$\pm$0.5 & $2.47\times 10^{-9}$ & 0.71 & 2.72 & 14.6 & \textcolor{green}{$\blacksquare$} \\
PVS$_8$ & $45\pm4$ & 89$\pm$0.5 & $2.04\times10^{-9}$ & 0.67 & 2.72 & 16.1 & \textcolor{green}{$\blacksquare$} \\
PVS$_8$ & $38\pm3$ & 89$\pm$0.5 & $1.23\times 10^{-9}$ & 0.59 & 2.72 & 20.1  & \textcolor{green}{$\blacksquare$} \\
PVS$_8$ & $32\pm3$  & 89$\pm$0.5 & $7.32\times 10^{-10}$ & 0.52 & 2.72 & 26.87 & \textcolor{green}{$\blacksquare$} \\
\hline
\end{tabular}
\caption[Details of thin elastic sheet properties]{Details of the properties of the elastic sheets used in the experiments concerning the onset of wrinkling, including: the material, the sheet thickness $h$, diameter $D$, bending stiffness $B$, the elasto-gravity bending length $\lg$, the liquid-vapour capillary length $\lc$ and $\tau = \lc^2 /\lg^2$. The table also includes the data marker used to denote the elastic sheet in figure~\ref{fig:Onset}. }
\label{table2}
\end{center}
\end{table}

\subsection{Linear stability and experimental results}


\begin{figure}
	\centering
\includegraphics[width=\textwidth]{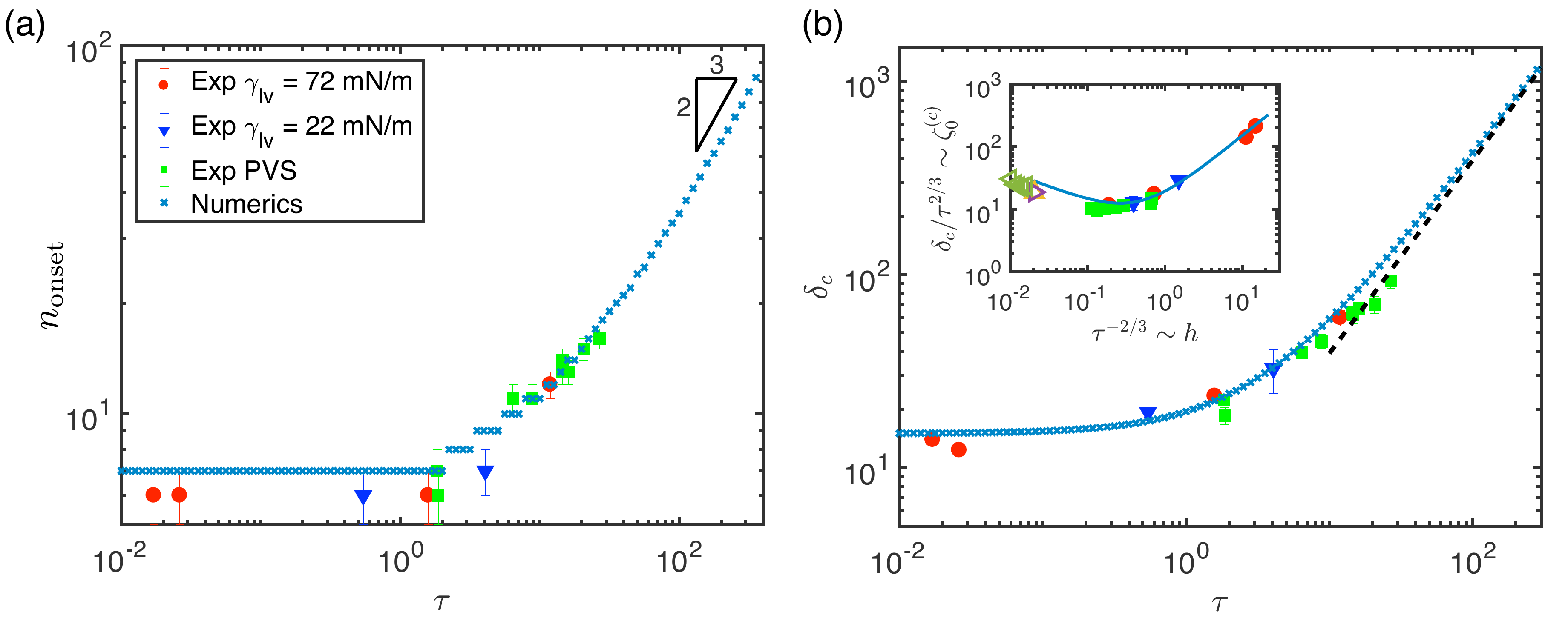}
	\caption{(a) The number of wrinkles observed at onset, $\nonset$ as a function of $\tau$.  (b) The dimensionless indentation depth at which a wrinkled solution first exists, $\woc$, and (inset) a measure of the dimensional depth at which wrinkling should occur as the thickness $h$ of the membrane is varied (see text). In both figures the markers denote the results of simulations ($\textcolor{pink2}{\times}$), the experiments performed on PDMS, PI and PC, for which $\glv=72\mathrm{~mN/m}$ ($\red{\bullet}$) and $\glv=22\mathrm{~mN/m}$ ($\blue{\blacktriangledown}$), and on PVS, for which $\glv=72\mathrm{~mN/m}$ ($\green{\blacksquare}$).  In (a) the triangle indicates the near-threshold scaling prediction, following \cite{Davidovitch2011}, for the wrinkle number, $n_c \sim \tau^{2/3}$. In (b) the  prediction of membrane theory, \eqref{eqn:largetauOnset}, valid for $\tau\gg1$ \cite{Vella2015}, is shown by the dashed line. The additional points in the inset (\textcolor{green2}{$\triangleleft$}, \small{\textcolor{orange}{$\triangle$}}, \textcolor{pink4}{$\triangleright$}) are taken from previous experiments with highly bendable sheets and $\nu=0.3 $\cite{Vella2015} illustrating the non-monotonicity of $\zeta_0^{(c)}$ as $h$ varies. We attribute the quantitative disagreement of these experiments to the finite size of the sheets used previously \cite{Vella2015}, though the different Poisson ratio may also make a small difference. In both figures,  error bars represent the standard deviation from at least 10 measurements.	}
	\label{fig:Onset}
\end{figure}

The key quantities of interest are the critical depth, $\woc$, at which wrinkling begins and $\nonset$, the number of wrinkles observed at this onset. Figure~\ref{fig:Onset}(a) and (b) therefore show how $\nonset$ and $\woc$, respectively, vary with the dimensionless mechanical bendability  $\tau$.  The experimental observations of both the number of wrinkles at onset and the critical indentation depth broadly agree with the results of the linear stability analysis.  In particular, both theory and experiment illustrate that the number of wrinkles and onset indentation are approximately constant for small mechanical bendability, $\tau \ll 1$. This is as should be expected: recall that the limit $\tau=0$ is regular and so, for $\tau\ll1$, the problem is governed by the indentation depth $\wo$ alone (or equivalently, the geometrical bendability $\epsG^{-1}=\wo^2$). This implies that the wrinkling instability must occur at some critical value $\woc$, independently of $\tau$. This critical value $\woc$ must therefore be some $O(1)$ constant for $\tau \ll 1$ (in agreement with the numerical and experimental results).

In the limit of large mechanical bendability, $\tau\gg1$,  we find that the numerical results  for the indentation depth at onset, $\woc$,  are consistent with previous results for $\tau=\infty$ \cite{Vella2015},  which is written in our dimensionless notation in \eqref{eqn:largetauOnset}, and is shown in figure \ref{fig:Onset}(b) for comparison. In this limit our numerical results for the wrinkle number suggest $\nonset\sim\epsM^{-1/3}\sim\tau^{2/3}$, which is the scaling expected by modifying a previous scaling analysis of the problem close to threshold \cite{Davidovitch2011}. Our experimental results reproduce the numerically expected values (to within experimental error) but do not reach the very large mechanical bendability regime where a true scaling exists. Instead, the experimental results taken on their own appear to suggest a scaling $\nonset\sim\tau^{1/2}$. We emphasize that this apparent scaling law is simply the transition between different scaling laws ($\nonset=O(1)$ at $\tau\ll1$ and $\nonset\sim \tau^{2/3}$ at $\tau\gg1$) and should not be relied upon. This is similar to the behaviour observed in a related, Near Threshold, buckling problem \cite{Taffetani2017}. Finally, we note that since $\woc/\tau\sim \tau^{-1/3}\to0$ as $\tau\to\infty$ the rescaling of the axisymmetric problem suggested in \eqref{eqn:rescale} is unphysical for sheets with high mechanical bendability, $\tau\gg1$: such sheets will wrinkle before they reach the limit $\wo/\tau\gg1$ for which the  axisymmetric state becomes universal and the length scale $\lcurv\propto\wo^{1/2}$ emerges, see figure \ref{fig:StressProfiles}(d). Whether similar scalings persist `Far from threshold' remains to be seen.

For many applications it may also be interesting to understand how the indentation depth required for wrinkling varies with sheet thickness as the other, material, properties of the system are maintained. In particular, with fixed $E$, $\rho g$ and $\glv$, the dimensional indentation depth at the onset of wrinkling has a minimum as the sheet thickness $h$ varies. To see this, consider first a scaling point of view:  one expects $\zeta_0^{(c)}\propto h$ for $\tau\ll1$, while previous work \cite{Vella2015} showed that $\zeta_0^{(c)}\propto h^{-1/2}$ in the limit  $\tau\gg1$. These qualitatively different scalings of $\zeta_0^{(c)}$ with $h$, combined with the monotonic decrease of $\tau$ with $h$ ($\tau\propto h^{-3/2}$ from \eqref{eqn:tauDefn}) leads us to expect that $\zeta_0^{(c)}$ will attain a minimum value at intermediate thicknesses. In more detail,  the dimensional indentation depth at the onset of wrinkling may be written (with other material properties assumed fixed) as a function of $\tau$ alone:
\beq
\zeta_0^{(c)}=h\times\woc(\tau)\sim \woc(\tau)/\tau^{2/3}.
\eeq  Our numerical results confirm the expectation that this quantity is minimized as the sheet thickness varies (see inset of figure~\ref{fig:Onset}(b)) with the minimum  located at $\woc/\tau^{2/3} \approx 12.6$ and $\tau^{-2/3} \approx 0.27$ for $\nu=0.5$. Our experiments were not able to reproduce this minimum cleanly but, when combined with previously published experimental data \cite{Vella2015}, do show a clear minimum.

The appearance of a minimum in the indentation depth required for wrinkling is surprising at first, but may be understood qualitatively by recalling that wrinkling requires both a sufficient level of compression and a sufficiently low bending rigidity. Thin sheets have a high mechanical bendability ($\tau \gg 1$) and so the precise value of their bending rigidity is irrelevant. Instead, the threshold for wrinkling is  governed by the compression level: wrinkling occurs when the azimuthal compression induced by indentation overcomes the applied (interfacial) tension. As a result, when the   tensile load is reduced (e.g.~by using a thicker, but still `thin', sheet) a lower compression level is needed to induce wrinkling. At the other end of the spectrum, thick sheets have a low mechanical bendability ($\tau \ll 1$) and hence can withstand large compressive forces before buckling through wrinkling. As a result, the threshold for wrinkling of such sheets is governed by the bending rigidity: the thinner a `thick' sheet is, the lower the compression level required for wrinkling. Between these two extremes, the indentation depth required for wrinkling is minimized.


\section{Discussion and Conclusions}

In this paper, we have investigated the response of a floating, elastic sheet to an applied, localized load focussing on the limit of low-to-moderate mechanical bendability, $10^{-2} \lesssim \tau \lesssim 10^{2}$. For loads insufficient to wrinkle the sheet, the resultant deformation is axisymmetric and is characterized by two regimes in the force-displacement law: with small displacements, the force is linearly proportional to the imposed indentation, while for large displacements the force is proportional to the square of the imposed indentation. These different responses can be understood as the result of the elastic object deforming over a horizontal length scale $\ell$ that is deformation-independent for small indentation depths (but varies with the sheet thickness) but that is deformation dependent for large indentation depths. For small indentation depths, the gravitational potential energy of the liquid displaced by this deformation $\sim \rho g\ell^2\zeta_0^2$, which balances the work done by the indenter, $\sim F\zeta_0$, giving a linear force--displacement relation. For large indentation depths, the deflection of the elastic object occurs over the horizontal length scale $\ell_\ast\sim\zeta_0^{1/2}$ and so the gravitational potential energy of the liquid displaced by this deformation $\sim \rho g\zeta_0^3$. The resulting force law is therefore quadratic in $\zeta_0$ (though we emphasize that the development of wrinkling far beyond threshold may return the system to a linear force law \cite{Vella2015}).

We began this study by posing the question of which material  we feel when we poke a sheet that coats a liquid layer. We can now answer this question, assuming  our poking is limited to small vertical displacements, by examining the small indentation spring stiffness
\beq
K_1=F/\zeta_0\sim\begin{cases}
8(B\rho g)^{1/2},\quad \tau\ll1,\\
\frac{2\pi}{\log(4\tau)} \glv,\quad \tau\gg1.
\end{cases}
\eeq With this result we see that for sheets with low mechanical bendability ($\tau\ll1$), the linear stiffness is a mixture of that due to the substrate and that due to the sheet itself.  This observation can be rationalized by the fact that the limit of zero mechanical bendability, $\tau=0$ is regular, and determined entirely between the balance between the bending stiffness of the sheet and the hydrostatic pressure within the liquid. However, for   sheets with high mechanical bendability ($\tau\gg1$) this stiffness is instead dominated by the surface tension of the interface with the mechanical properties of the sheet entering only via a logarithmic correction. In the analogous case of a Winkler foundation --- an elastic composite comprising a thin sheet bonded to a substrate that provides a linear restoring force --- we anticipate that a similar result will hold: for relatively unbendable sheets the stiffness will result from a combination of the sheet and the substrate, while for highly bendable sheets the stiffness will be dominated by any boundary tension existing in the sheet prior to indentation.

Larger indentations cause a significant perturbation to the stress (compared to the stress prior to  indentation). This is the origin of the emergent length scale $\ell_\ast$, and the transition between the small-displacement and large-displacement force laws. However, another consequence of the change in the stress is that the hoop stress becomes compressive: material is pulled to a radial position at which its natural length is too great. If this compression becomes large enough, the sheet relieves this frustration by buckling out of the plane, with radial wrinkles forming. We have analysed the onset of this wrinkling instability as a function of the mechanical bendability $\tau$ determining both the critical indentation depth required to bring about wrinkling, and the number of wrinkles observed at onset. Our result for the critical indentation required to generate wrinkling in the high mechanical bendability, $\tau\gg1$, limit agree with those determined from membrane theory and verified experimentally previously \cite{Vella2015}. Our results for the number of wrinkles at onset in this regime suggest that $\nonset\sim\tau^{2/3}$ for highly bendable sheets. Our study of the onset of wrinkling at low and moderate mechanical bendabilities, $\tau\lesssim1$, suggests that both the onset indentation, and the number of wrinkles at onset are $O(1)$ quantities as $\tau\to0$, and are  confirmed by our own experiments. 
 
We note in closing that our analysis of the onset of wrinkling assumed that wrinkling occurs with a single wavenumber. In the case of well-developed wrinkling (far-from-threshold) with high bendability, it has recently been observed that the wrinkle number  may, in fact, evolve with both indentation depth and radial position \cite{Paulsen2016}. More detailed analysis of this problem beyond onset is needed to understand whether this spatial variation is also observed at low and moderate mechanical bendabilities --- the parameter regime highlighted as `open' in table \ref{table:DG}. Nevertheless, our study of the small-indentation behaviour up to and including the threshold of instability extends our understanding of deformation in such scenarios. In particular, our exploration of the role of the dimensionless parameters $\tau$ and $\wo$ in this problem allows the behaviour of a wide range of systems from the nano-indentation of ultra-thin films to the geological context of the loading of ice sheets and tectonic plates to be studied within a single framework. \\

\section*{Acknowledgements}

We are grateful to  Benny Davidovitch for discussions of this work. The research leading to these results has received funding from the European Research Council under the European Union's Horizon 2020 Programme / ERC Grant Agreement no.~637334 (DV).

%

\end{document}